\documentclass[sigconf]{acmart}

\copyrightyear{2026}
\acmYear{2026}
\setcopyright{cc}
\setcctype{by}
\acmConference[ICAIF '26]{7th ACM International Conference on AI in Finance}{November 14--17, 2026}{Milan, AA, Italy}
\acmBooktitle{7th ACM International Conference on AI in Finance (ICAIF '26), November 14--17, 2026, Milan, AA, Italy}
\acmDOI{10.1145/3858063.3858783}
\acmISBN{979-8-4007-3087-0/2026/11}

\usepackage{booktabs}
\usepackage{graphicx}
\usepackage{amsmath}
\usepackage{multirow}
\usepackage{algorithm}
\usepackage{algpseudocode}
\usepackage{textcomp}   
\usepackage{setspace}
\newcommand{\eur}{\texteuro}
\begin{document}

\title{One Basket, Many Experts: Decorrelated Re-ranking \\ for Financial Asset Recommendation}

\author{Ryuki Unno}
\affiliation{%
  \institution{Hokkaido University}
  \city{Sapporo, Hokkaido}
  \country{Japan}
}
\email{unno@lmd.ist.hokudai.ac.jp}

\author{Keigo Sakurai}
\affiliation{%
  \institution{Hokkaido University}
  \city{Sapporo, Hokkaido}
  \country{Japan}
}
\email{sakurai@lmd.ist.hokudai.ac.jp}

\author{Koshi Watanabe}
\affiliation{%
  \institution{Hokkaido University}
  \city{Sapporo, Hokkaido}
  \country{Japan}
}
\email{koshi@lmd.ist.hokudai.ac.jp}

\author{Keisuke Maeda}
\affiliation{%
  \institution{Hokkaido University}
  \city{Sapporo, Hokkaido}
  \country{Japan}
}
\email{maeda@lmd.ist.hokudai.ac.jp}

\author{Takahiro Ogawa}
\affiliation{%
  \institution{Hokkaido University}
  \city{Sapporo, Hokkaido}
  \country{Japan}
}
\email{ogawa@lmd.ist.hokudai.ac.jp}

\author{Miki Haseyama}
\affiliation{%
  \institution{Hokkaido University}
  \city{Sapporo, Hokkaido}
  \country{Japan}
}
\email{mhaseyama@lmd.ist.hokudai.ac.jp}

\renewcommand{\shortauthors}{Unno et al.}

\begin{abstract}
Financial asset recommendation for retail investors is typically optimized for relevance, that is, how well a recommendation matches what each user will buy.
From the service provider's point of view, the value of a recommendation also depends on portfolio performance: the risk-adjusted return the recommended assets (the basket) deliver, which drives user engagement and fee revenue. 
The two goals are rarely achieved jointly, because the recommendations that best match user behavior concentrate on a few popular assets whose returns are highly correlated, which caps the risk-adjusted return a provider can offer.
We propose Decorrelated-Satellite Expert Re-ranking (\textbf{DSER}), which retains the most relevant recommendations from an expert-following list and replaces the rest with return-decorrelated alternatives. 
The number of replaced slots is a single integer~$m$ (a knob) that the provider sets, so the relevance--performance trade-off becomes an explicit, auditable choice rather than a property hidden inside a learned model.
In experiments on the public FAR-Trans retail-brokerage dataset, under the primary protocol, replacing a few slots improves risk-adjusted return at a small relevance cost, and the gain survives realistic trading costs at typical retail portfolio sizes.
Minimizing correlation would lower basket volatility, but in our data the least-correlated assets are more volatile and higher-return, so their volatility offsets that reduction while their return drives the gain, not lower volatility.
The resulting trade-off is asymmetric, with the relevance cost consistent across market phases while the gain is phase-dependent. The provider sets $m$ to read off exactly how much relevance each operating point costs.
\end{abstract}



\maketitle

\section{Introduction}
\label{sec:intro}

\emph{Financial Asset Recommendation} (FAR)~\cite{zibriczky2016} is now a well-established task in both practice and research.
Banks, brokers, and robo-advisors deploy it as a routine service, and researchers study it as a recommender-systems task over transaction histories, with a public multi-algorithm benchmark~\cite{fartrans} and methods built on real brokerage data~\cite{pfotgnrec}.
It guides retail investors who must choose among thousands of assets under a flood of information. Overwhelmed, many settle for a handful of popular names and hold poorly diversified portfolios.

FAR is set apart from ordinary recommendation since a useful recommendation must succeed on two distinct axes.
The first is \textbf{recommendation relevance}: how well the ranked list matches the assets a user actually goes on to buy (scored per user by nDCG and hit rate), which sustains the engagement and retention a provider ultimately depends on~\cite{jannach2019}.
The second is \textbf{portfolio performance}: the return (ROI) and risk-adjusted return (Sharpe ratio) of the recommended list when held as an equal-weight basket.
That benchmark of eleven FAR algorithms found no method superior on both axes~\cite{sanzcruzado2022,fartrans}.
Optimizing purely for relevance leaves returns near the market average, while optimizing purely for return ignores what users will actually adopt~\cite{mvecf}.

A strong, interpretable baseline is expert following~\cite{expertfollow}, which ranks assets by what high-performing investors hold and fits no parameters.
Its appeal is that this single signal speaks to both axes at once.
On the performance side, the past returns of top investors tend to persist~\cite{barber2014skill,coval2021individual}, so the experts' holdings remain competitive in risk-adjusted terms~\cite{expertfollow}.
On the relevance side, experts are themselves retail investors, so their most-held assets overlap with what typical users go on to buy.

The expert-following list's weakness is \emph{crowding}.
Because expert consensus concentrates on a few popular, heavily traded assets, the top-$k$ list has a high average pairwise return correlation. 
Its assets move together, diversify poorly, and are exposed to common shocks~\cite{markowitz1952,rura2026}, which caps the list's risk-adjusted return.
Lowering that correlation creates what we call a diversification budget. A less correlated basket is less volatile than its individual assets, so there is room to admit higher-volatility assets without raising the basket's overall risk.
In our data this budget is spent on return, not on lower volatility, since the least-correlated assets are higher-return (Section~\ref{sec:ablation}).
This motivates raising the risk-adjusted return of an expert-following list by replacing only its lowest-ranked items with decorrelated assets, without materially sacrificing relevance.

\begin{figure*}[!t]
  \centering
  \includegraphics[width=0.78\textwidth]{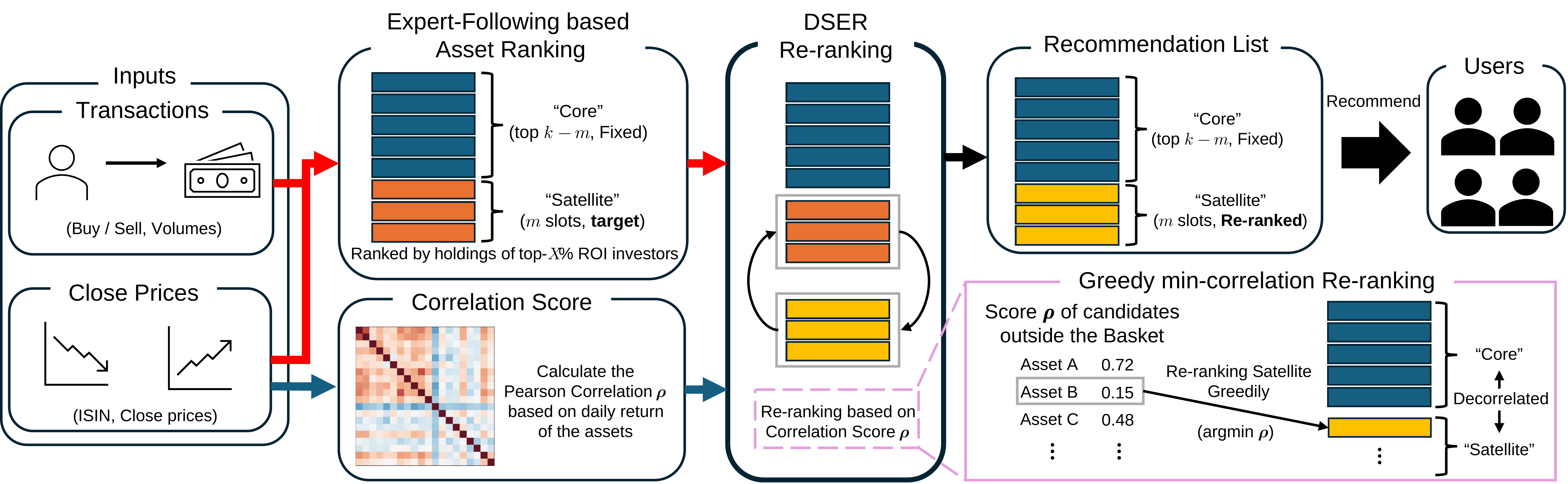}
  \Description{Pipeline diagram. Transaction records and daily close prices feed two modules: an expert-following ranking of assets, split into a fixed core of the top $k-m$ items and $m$ satellite slots, and a Pearson correlation matrix of daily returns. DSER re-ranks only the satellite slots, greedily adding the candidate with the lowest correlation to the current basket, and the resulting list of core and decorrelated satellites is recommended to all users.}
  \caption{The DSER pipeline. An expert-following ranking is split into a fixed
  relevance core (the top $k-m$ items) and $m$ satellite slots. DSER re-ranks
  only the satellites, filling each slot with the asset of minimal correlation
  to the current basket ($\arg\min\rho$). The knob $m$ sets how much of the list is decorrelated.}
  \label{fig:pipeline}
\end{figure*}

We propose Decorrelated-Satellite Expert Re-ranking (\textbf{DSER}), a plug-in re-ranker that controls the trade-off between the two objectives. 
It keeps the expert signal as the recommendation's relevance backbone while explicitly decorrelating its constituents (Figure~\ref{fig:pipeline}). 
We frame it as a transparent control and characterize it on the curated, liquid universe a provider actually operates in.
Concretely, it retains the top $k-m$ expert items as a relevance core and fills the remaining $m$ satellite slots with the assets whose mean correlation to those already chosen is lowest.
For instance, at $k{=}10$ and $m{=}2$ it keeps the eight strongest expert picks and replaces the last two slots with the least-correlated available assets.
The single integer $m$ is the provider's only control, set rather than learned.
DSER's criterion is the assets' full return-correlation structure, not the diagonal variance penalty of the nearest re-ranker RURA~\cite{rura2026}, which DSER outperforms on the combined score (Section~\ref{sec:main}).

In summary, our contributions are threefold.
\begin{enumerate}
\item[(i)] \textbf{A training-free, auditable re-ranker}: DSER, a core--satellite method that turns the relevance--performance trade-off into a single integer knob $m$, with each substitution declared, not learned (Section~\ref{sec:method}).
\item[(ii)] \textbf{A joint evaluation} of recommendation relevance and portfolio performance on FAR-Trans, showing that the knob raises risk-adjusted return at an explicit relevance cost, a gain that survives realistic trading costs at typical retail portfolio sizes (Sections~\ref{sec:main},~\ref{sec:capacity}).
\item[(iii)] \textbf{A mechanism and robustness analysis} explaining why it works, attributing the gain to a volatility-matched return improvement (decorrelation admits higher-return satellites at flat basket volatility in the deployable band) and bounding its magnitude under stricter protocols (Sections~\ref{sec:ablation},~\ref{sec:robust}).
\end{enumerate}

\section{Related Work}
\label{sec:related}

\paragraph{Financial Asset Recommendation.}
As a recommendation task~\cite{zibriczky2016}, FAR divides precisely along the axis this paper targets.
Relevance-oriented methods predict what a user will buy, including news- and sentiment-based learning-to-rank~\cite{song2017}, attention- and personality-aware recommenders~\cite{takayanagi2023,takayanagi2024}, and collaborative filtering for assets~\cite{zheng2020}. 
Return-oriented methods instead optimize financial outcomes, such as large-scale portfolio-optimization recommendation~\cite{swezey2018}.
A benchmark on FAR-Trans shows that transaction-based relevance is only a proxy for portfolio performance and that no method dominates on both axes~\cite{sanzcruzado2022,fartrans}. 
We adopt FAR-Trans, the first public dataset with an eleven-algorithm benchmark, and address this tension directly.
A related family builds the risk objective into the model itself, on standard architectures such as pairwise ranking (BPR~\cite{bpr}), graph collaborative filtering (LightGCN~\cite{lightgcn} and self-supervised variants~\cite{sgl}), and temporal graph networks over the transaction stream~\cite{tgn,tgat}.
The most direct precursor, PfoTGNRec~\cite{pfotgnrec}, couples a temporal graph recommender with mean-variance-aware sampling on FAR-Trans. 
MVECF~\cite{mvecf} folds mean-variance efficiency into the collaborative-filtering loss.
These models embed diversification in learned weights and expose no explicit control over it.
In contrast, DSER externalizes the trade-off as a thin, model-agnostic re-ranking layer that in principle sits on top of any such ranker.

\paragraph{Diversification and Re-ranking.}
Holding weakly correlated assets is the foundation of Modern Portfolio Theory~\cite{markowitz1952}, and the Sharpe ratio~\cite{sharpe1966} is the standard measure of the risk-adjusted return that theory targets. 
Estimates of a full covariance matrix are unstable in high dimensions, motivating shrinkage estimators~\cite{ledoitwolf2004}. 
The equal-weight ($1/N$) portfolio is famously hard to beat once estimation error is accounted for~\cite{demiguel2009}, which is why our baskets are equal-weighted.
Re-ranking for diversity is itself well studied in general recommendation, including maximal marginal relevance (MMR)~\cite{carbonell1998}, topic diversification~\cite{ziegler2005}, and determinantal point processes~\cite{chen2018dpp}. 
Yet these notions of diversity are content- or category-based rather than based on return covariance.
In finance, the closest work is RURA~\cite{rura2026}, which re-ranks upstream candidates by a risk-aware utility with a variance-based (diagonal) penalty and a single diversification knob~$\gamma$.
Our contribution is the financial diversity criterion, not the re-ranking framework. 
Sections~\ref{sec:main} and~\ref{sec:ablation} show that our integer knob sacrifices nothing relative to a $\lambda$-swept MMR in the operating region providers use, and that the return-correlation criterion outperforms a variance-only rule, including a reimplementation of RURA following its published description. 
This is consistent with the low-correlation tail carrying a return premium that a variance-only rule forgoes.

\paragraph{Expert Following.}
Following the trades of top-performing investors is a strong, interpretable, and competitive baseline on FAR-Trans~\cite{fartrans,expertfollow}.
Its weakness is precisely crowding. The expert consensus concentrates on a few popular assets, so the top-$k$ basket is internally highly correlated. 
DSER relaxes this structure while keeping the expert signal as the relevance backbone.
One caution is specific to this setting. 
Stale prices and return smoothing~\cite{lo2002,getmansky2004} depress a thinly traded asset's measured variance and correlation, mechanically inflating its realized Sharpe ratio. 
This threatens any correlation-based re-ranker, which may ``discover'' spuriously decorrelated stale assets.
We therefore restrict the recommendable universe to a curated, liquid asset pool, as prior work does~\cite{pfotgnrec}.
We then report the after-cost capacity of the (less-liquid) satellites under a liquidity-based impact model (Section~\ref{sec:capacity}).

\section{Method: DSER as a Provider Knob}
\label{sec:method}

\paragraph{Problem Setting.}
At each decision time, a provider must issue a size-$k$ ranked list that balances the two objectives from Section~\ref{sec:intro}. 
Relevance is scored against each user's subsequent purchases, and portfolio performance is the risk-adjusted return of the list held as an equal-weight basket. 
DSER takes the provider's expert-following ranking as input and re-ranks it into this list.

We design DSER around the provider's problem rather than a model class.
Because the expert list's crowding stems from which assets the consensus holds (Section~\ref{sec:intro}), the remedy acts on selection rather than weights, lowering that correlation while largely preserving relevance and remaining training-free, through a single core--satellite step with one integer knob $m$ (Figure~\ref{fig:pipeline}).

\subsection{Setup: Input and Baseline}
At each decision time $t$ the provider has computed an expert-following score over all tradeable assets~\cite{expertfollow}.
We identify the top $X\%$ of investors by trailing ROI (the \emph{experts}) and score each asset $i$ by their ROI-weighted purchases,
\begin{equation}
\label{eq:expert}
\mathrm{score}(i)\;=\;\sum_{u\in\text{top-}X\%}\max(\mathrm{ROI}_u,0)\cdot H_{u,i},
\end{equation}
where $H_{u,i}$ is the number of times user $u$ bought asset $i$ in the transaction window used to select the experts, and $\max(\mathrm{ROI}_u,0)$ weights each expert by their ROI while discarding loss-makers. 
Sorting assets by this score gives the full ranking $E_t=(e_1,e_2,\dots)$.
Its top-$k$ prefix $E^{(k)}_t=(e_1,\dots,e_k)$ is the unmodified expert-following baseline.
Because $E^{(k)}_t$ concentrates on popular assets, its average pairwise return correlation is high.
DSER takes the ranking $E_t$ and returns a re-ordered size-$k$ list that lowers this internal correlation while retaining the highest-scoring expert items.

\subsection{Core--Satellite Re-ranking}
DSER takes a single integer hyper-parameter $m\in\{0,1,\dots,k{-}1\}$, so at least one relevance item is always kept.
It keeps the top $k-m$ expert items as an ordered relevance \textbf{core} $C=(e_1,\dots,e_{k-m})$ and appends $m$ \textbf{satellites} one at a time. 
We measure co-movement by the Pearson correlation of two daily-return series $x,y$ over the trailing window,
\begin{equation}
\label{eq:pearson}
\rho(x,y)=\frac{\sum_{\tau}(x_\tau-\bar x)(y_\tau-\bar y)}{\sqrt{\sum_{\tau}(x_\tau-\bar x)^2}\,\sqrt{\sum_{\tau}(y_\tau-\bar y)^2}} .
\end{equation}
Each satellite is then the not-yet-chosen universe asset with the smallest mean correlation to the current basket:
\begin{equation}
\label{eq:select}
s \;=\; \arg\min_{a \notin C \cup S}\;
\frac{1}{|C\cup S|}\sum_{j\in C\cup S} \rho(r_a, r_j),
\end{equation}
where $r_a$ is the daily-return series of candidate asset $a$.
Satellites are appended in selection order, forming $S=(s_1,\dots,s_m)$, until the list holds $k$ assets.
The rule is greedy and sequential, each satellite chosen against the basket built so far, so the $\arg\min$ in Eq.~\ref{eq:select} is re-evaluated after every addition. Because $m$ is a fixed budget, the loop always fills $m$ slots, returning the least-correlated available asset even when that asset is not uncorrelated in absolute terms, which is why pushing $m$ toward the fully decorrelated end yields diminishing, then degenerate, relevance (Section~\ref{sec:main}).
The ordered output is $(e_1,\dots,e_{k-m},\allowbreak s_1,\dots,s_m)$.
Both inputs use only data available at $t$. The expert scores (Eq.~\ref{eq:expert}) use the transactions from the start of the current walk-forward instance up to $t$ (Section~\ref{sec:setup}), and the correlations (Eq.~\ref{eq:pearson}) use the daily returns of the past $180$ days.
Algorithm~\ref{alg:dser} states the procedure.

\begin{algorithm}[t]
\caption{DSER re-ranking at decision time $t$}
\label{alg:dser}
\begin{algorithmic}[1]
\Require expert ranking $E_t=(e_1,e_2,\dots)$; list size $k$; knob $m$; trailing returns $\{r_a\}$ over the $180$-day window ending at $t$
\State $C \gets (e_1,\dots,e_{k-m})$ \Comment{ordered relevance core}
\State $S \gets (\,)$ \Comment{ordered satellite sequence}
\While{$|S| < m$}
  \State $s \gets \arg\min_{a\notin C\cup S}\ \tfrac{1}{|C\cup S|}\sum_{j\in C\cup S}\rho(r_a,r_j)$ \Comment{least correlated to basket}
  \State append $s$ to $S$
\EndWhile
\State \Return ordered list $(C, S)$
\end{algorithmic}
\end{algorithm}

\paragraph{Design Choices.}
Three choices keep the rule simple and auditable.
(1) \emph{Correlation, not variance.} A rule that ranks assets by their own volatility ignores cross-asset covariance structure. Minimizing mean correlation targets that structure, which is both more effective (Section~\ref{sec:ablation}) and more interpretable than a variance-only rule.
(2) \emph{Greedy, basket-relative selection.} Each satellite is scored against the basket built so far (Eq.~\ref{eq:select}), so the chosen assets decorrelate jointly rather than individually. This avoids appending $m$ mutually correlated ``diversifiers.''
(3) \emph{Exact core preservation.} The top $k-m$ relevance items are never touched, which bounds the relevance loss and keeps the most liquid, most-traded consensus assets in the basket. Only the appended satellites are comparatively less liquid.

By construction, $m{=}0$ recovers the unmodified expert list. Empirically, larger $m$ monotonically decorrelates the basket and lowers relevance, so $m$ is a single, ordered control on the relevance--performance trade-off.
Because each satellite is a named, correlation-justified substitution and nothing is learned, the provider chooses and justifies the operating point.
\section{Experiments}
\label{sec:experiments}

We organize our study around four questions.
\begin{enumerate}
\item[\textbf{RQ1.}] Does increasing $m$ improve the portfolio performance a provider can deliver, and at what cost in relevance (Section~\ref{sec:main})?
\item[\textbf{RQ2.}] Does the gain survive realistic trading costs at retail portfolio sizes (Section~\ref{sec:capacity})?
\item[\textbf{RQ3.}] What inside DSER drives the gain (Section~\ref{sec:ablation})?
\item[\textbf{RQ4.}] Does the gain hold under stricter evaluation protocols (Section~\ref{sec:robust})?
\end{enumerate}

\subsection{Experimental Setup}
\label{sec:setup}

\paragraph{Data and Universe.}
We use FAR-Trans~\cite{fartrans}, a public dataset of real retail brokerage transactions and daily close prices spanning January 2018 to November 2022 ($29{,}090$ customers, $388{,}049$ transactions, and $321$ traded assets).
To our knowledge, it is the only public FAR dataset pairing real brokerage transactions with prices~\cite{fartrans}, so we evaluate on it alone, as is standard in this line of work.
Trading is concentrated. The $12$ most-traded assets account for over half of all transactions.
Following PfoTGNRec~\cite{pfotgnrec}, we build \textbf{U92} from the $321$ FAR-Trans assets by keeping those retrievable from public price feeds and dropping any asset whose price is unchanged for $30$ consecutive days, leaving $92$ exchange-listed equities.
This is an absolute tradeability bar, not a claim that the $92$ assets are equally liquid. It removes assets untradeable at retail scale, whose stale prices would also let a correlation-based rule manufacture spurious diversification (Section~\ref{sec:related}). Relative liquidity still varies within U92. The popular core is heavily traded, while the appended satellites come from its thinner but still tradeable tail (Section~\ref{sec:capacity}).
This curation follows prior work~\cite{pfotgnrec} and mirrors the preprocessing a provider performs in practice.
Although U92 covers only $\sim$29\% of traded assets, it carries $\sim$62\% of all buy orders, so it is where a provider's recommendations actually land. Because DSER is a construction over expert scores and return correlations rather than a fitted model, its operating-point framework does not depend on this particular universe, even though the specific trade-off magnitudes and the reliability of the correlation estimates do.

\paragraph{Metrics.}
For each recommended equal-weight basket $B$ held over the evaluation window, we report return and risk,
\begin{equation}
\mathrm{ROI@}k = \tfrac{1}{k}\!\sum_{i\in B}\!\frac{p_i^{\text{end}}-p_i^{\text{start}}}{p_i^{\text{start}}},
\qquad
\mathrm{SR@}k = \sqrt{252}\,\frac{\overline{r}_B}{\sigma_{r_B}},
\end{equation}
where $\overline{r}_B$ and $\sigma_{r_B}$ are the mean and standard deviation of the daily basket return, and we set $r_f{=}0$, following prior FAR-Trans evaluations~\cite{expertfollow,pfotgnrec}. The mean return and volatility we report are annualized (daily mean $\times 252$, daily standard deviation $\times\sqrt{252}$). ROI@$k$ is the unannualized holding-period return. SR@$k$ uses the daily return of the basket rebalanced to equal weights each day, whereas ROI@$k$ is the buy-and-hold return of an equal initial investment in each asset.
We complement these with two relevance metrics computed against users' actual subsequent purchases: Normalized Discounted Cumulative Gain (nDCG@$k$), the standard position-discounted ranking metric~\cite{jarvelin2002}, and Hit rate (HR@$k$), which indicates whether the user subsequently buys at least one of the recommended items.
DSER's purpose is to trade a controlled amount of relevance for portfolio performance.

\paragraph{Baselines.}
A fair comparison must cover both ends of the relevance\nobreakdash--performance axis (Section~\ref{sec:related}), so we draw the full breadth of the FAR-Trans benchmark~\cite{fartrans} and group it by design objective. \emph{Performance-based} baselines (the market-average heuristic and a return-oriented random-forest ranker) target returns. \emph{Relevance-based} baselines (popularity, the collaborative filters BPR~\cite{bpr}, LightGCN~\cite{lightgcn}, and SGL~\cite{sgl}, and the temporal graph recommenders TGN~\cite{tgn} and TGAT~\cite{tgat}) predict what users buy. \emph{Hybrid} baselines (the mean-variance filter MVECF~\cite{mvecf} and PfoTGNRec~\cite{pfotgnrec}) combine both. For MVECF, we train the authors' weighted-matrix-factorization model on users' purchases from the start of the training span up to each decision time, with $\lambda_{\mathrm{MV}}{=}10$ and $\gamma{=}5$, the setting with the largest Sharpe improvement in the original paper. The original estimates mean returns and covariances from five years of returns, longer than the price history available at any of our decision times, so we instead estimate them over the past $30$ days, the window PfoTGNRec uses for its mean--variance rating. We reimplement PfoTGNRec from its description~\cite{pfotgnrec}. It computes this rating from the $30$ days after each interaction, which are not available at a decision time, so our reimplementation uses the $30$ days before it.
All learned baselines are re-trained, and every baseline is re-scored, under our own walk-forward protocol on U92, rather than imported from published numbers, so every method sees the same instances.\footnote{Learned baselines use the settings recommended in their original papers ($64$-dimensional embeddings, Adam with a learning rate of $10^{-3}$, and $20$ epochs, with PfoTGNRec at its searched dropout of $0.2$ and $\lambda_{\mathrm{mv}}{=}0.1$). MVECF uses its own settings ($30$ factors, a confidence weight of $10$, and $20$ alternating-least-squares epochs). Code, the full baseline hyper-parameter grid, and per-instance results will be released upon publication. DSER's headline comparison is internal (against the expert list it re-ranks), so it does not hinge on baseline tuning.}
The expert-following list (the top $X\%$ of investors as experts, with $X{=}5$ and list size $k{=}10$) is the anchor that DSER re-ranks. We rank experts by trailing ROI rather than by a risk-adjusted score, a choice we ablate in Section~\ref{sec:ablation}.
We further compare against a reimplementation of RURA~\cite{rura2026}, the nearest prior-art re-ranker (Section~\ref{sec:related}), and a minimum-variance-satellite control that isolates what in DSER matters.
Like the expert anchor it re-ranks~\cite{expertfollow}, DSER issues one list for all users.

\begin{figure}[t]
  \centering
    \includegraphics[width=0.9\linewidth]{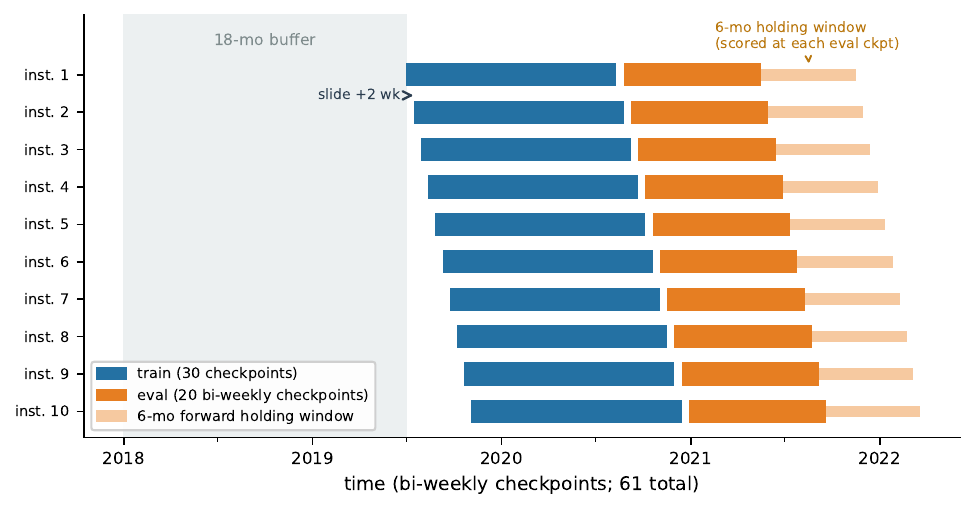}
    \Description{Timeline from 2018 to 2022 with ten horizontal bars, one per walk-forward instance. After an 18-month buffer, each instance has a 30-checkpoint training span followed by a 20-checkpoint evaluation span and a 6-month forward holding window. Each instance starts two weeks after the previous one, so the instances overlap heavily.}
  \caption{Walk-forward protocol. $61$ bi-weekly checkpoints follow an $18$-month buffer.
  Each instance is a $30$-checkpoint train span followed by a $20$-checkpoint eval span, with each
  eval point scored over a $6$-month forward window. 
  Sliding the block forward by $2$ weeks yields $12$ placements, the first $10$ of which we average over.}
  \label{fig:protocol}
\end{figure}

\paragraph{Protocol.}
We take the recommendable universe from PfoTGNRec~\cite{pfotgnrec} and otherwise follow the walk-forward protocol of the expert-following study~\cite{expertfollow}, which we spell out here (Figure~\ref{fig:protocol}) because it defines our $n{=}10$ sample.
After an $18$-month initial buffer (January 2018--June 2019), we place a checkpoint every two weeks up to October 2021, giving $61$ checkpoints.
A single walk-forward instance is $30$ training checkpoints followed by $20$ evaluation checkpoints. 
The training checkpoints exist only to fit the learned baselines. Training-free DSER fits nothing on them. It only reads the transactions from the start of this span up to each decision time $t$ to select the experts, and the past $180$ days of returns to compute the correlations. 
At each of the $20$ evaluation checkpoints $t$, we build the recommendation from the data available up to $t$ and score it against each user's realized purchases over the following $6$~months (the holding window over which ROI@$k$ and SR@$k$ are also measured).
Within an instance, each metric is averaged over the same active-user$\,\times\,$checkpoint observations across the $20$ decision points. 
The two families differ only in their unit of analysis. ROI@$k$ and SR@$k$ are basket-level (identical for the users who share the one list), whereas nDCG@$k$ and HR@$k$ are scored per user against realized purchases.
We then slide the whole block forward one checkpoint (two weeks) at a time. 
The $61$ checkpoints admit $12$ such placements, of which we take the first $10$, matching the $n{=}10$ of the expert-following protocol we adopt~\cite{expertfollow}. 
We report each metric as the mean across these $10$ instances. Because the $6$-month windows are long relative to the two-week stride, the instances overlap. 
We acknowledge this and re-test on genuinely non-overlapping windows in Section~\ref{sec:robust}.
The expert signal, correlations, and selection are recomputed at every checkpoint (omitting this recomputation degrades the expert signal).
We use the buy-more setting (already-held assets remain recommendable), matching the FAR-Trans convention that many purchases add to existing holdings.
Instances are deterministic, so method comparisons are paired. 
We lead with effect size, direction, and consistency across instances, and report significance tests where the sample supports them (Section~\ref{sec:robust}).

\subsection{Main Result: $m$ as a Tunable Trade-Off}
\label{sec:main}

RQ1 asks whether increasing $m$ raises the portfolio performance a provider can deliver, and at what cost in relevance.

\begin{figure}[t]
  \centering
  \includegraphics[width=0.93\linewidth]{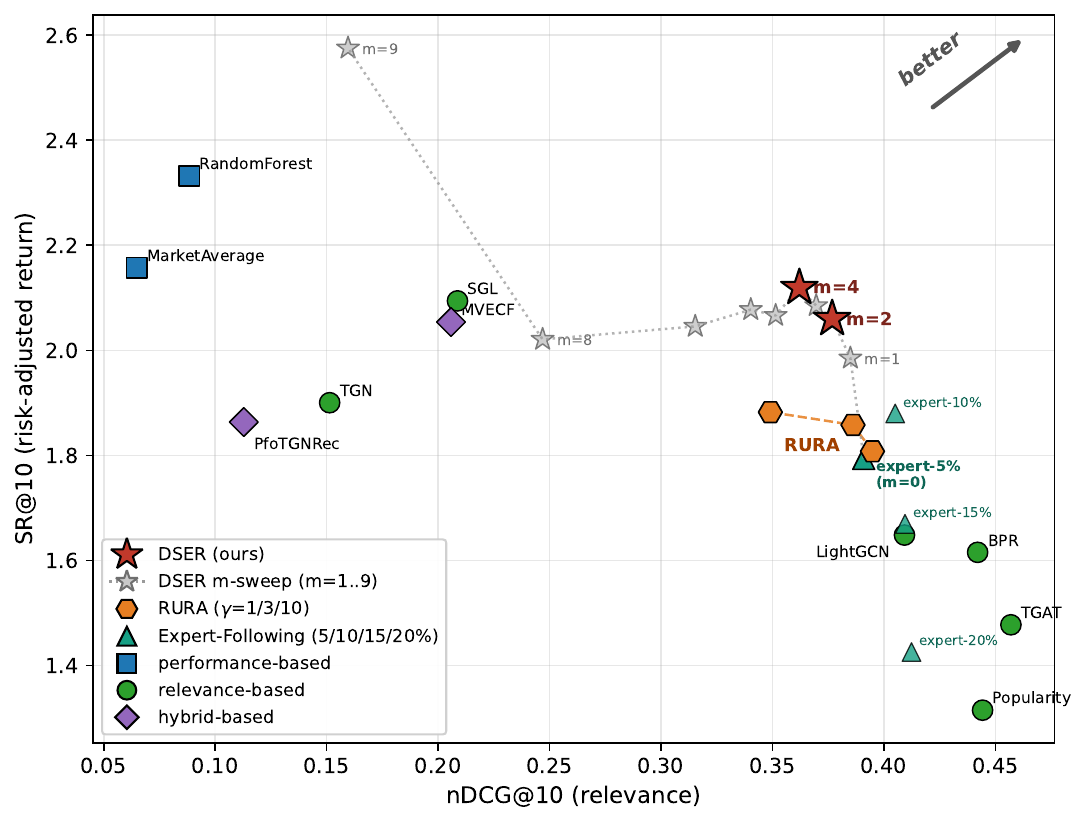}
  \Description{Scatter plot of SR@10 against nDCG@10 for all methods on U92. Performance-based baselines have high Sharpe ratios and near-zero relevance, relevance-based baselines have high relevance and lower Sharpe ratios, and the hybrid baselines have low relevance. The expert-following lists and RURA lie between these groups, around nDCG 0.39 to 0.41. The DSER m-sweep rises from the expert list to the m=2 and m=4 points (Sharpe about 2.1 at nDCG 0.36 to 0.38), then moves toward low relevance and ends at m=9 with Sharpe 2.58 and nDCG 0.16.}
  \caption{The relevance--performance landscape on U92 (walk-forward, $n{=}10$, $k{=}10$).
  Grey stars trace the $m$-sweep, red points mark DSER ($m{=}2,4$), and orange the
  RURA frontier ($\gamma$ swept), which DSER lies above across the deployable band.
  Baseline markers follow the design categories of Section~\ref{sec:setup}.}
  \label{fig:landscape}
\end{figure}

\paragraph{Expert following provides a competitive joint anchor.}
In Figure~\ref{fig:landscape}, every method with a clearly higher Sharpe ratio than the expert list (RandomForest, MarketAverage, SGL, MVECF, TGN, PfoTGNRec; SR $>1.86$) has substantially lower relevance (nDCG $\le 0.21$, and near zero for MarketAverage and RandomForest, against $0.391$ for the expert list), while every benchmark method that exceeds the expert list's relevance (Popularity, BPR, LightGCN, TGAT; nDCG $>0.40$) has substantially lower Sharpe.
Among the benchmark methods, only expert following occupies the high-relevance, competitive-Sharpe corner (nDCG $0.391$, SR $1.794$).
MVECF is hybrid by design, but at its most portfolio-efficient setting its mean-variance term steers the list away from the crowded consensus assets users buy, raising its Sharpe ratio ($2.054$) while its relevance falls to $0.206$. Design intent does not fix a method's position in this plane, as the relevance-based SGL and TGN, and the hybrid MVECF and PfoTGNRec, all land at low relevance.
The prior-art re-ranker RURA edges the anchor on both axes (nDCG $0.395$, SR $1.830$) and is compared below.
The provider's remaining question is whether this anchor's Sharpe ratio can be raised without leaving the high-relevance region.

On the curated universe U92, sweeping $m$ traces an empirical relevance--performance trade-off curve (Table~\ref{tab:u92}, Figure~\ref{fig:landscape}).
Starting from the unmodified expert list ($m{=}0$: SR $1.794$, nDCG $0.391$), Sharpe improves steeply through $m{\le}4$ (at only a small relevance cost) and turns non-monotonic thereafter.
At $m{=}2$, the provider gains SR $+0.27$ and ROI $+0.031$ at a relevance cost of $0.014$ nDCG ($-3.6\%$) and $0.033$ HR ($-4.5\%$). At $m{=}4$, SR rises by $0.33$ for a larger cost ($-0.029$ nDCG, $-0.065$ HR).
The intermediate $m{=}3$ is a balanced compromise, beating $m{=}2$ on Sharpe ($2.085$ vs.\ $2.060$) and $m{=}4$ on ROI ($0.266$ vs.\ $0.259$). We foreground $m{=}2$ and $m{=}4$ as the low-cost and Sharpe-peak points within the deployable band, but the provider reads the whole sweep of Table~\ref{tab:u92}, not a single $m$.
Both SR gains are positive in all $10$ paired instances (across-instance std $\le0.05$).
The sweep flattens across $m{=}5$--$8$, where the Sharpe ratio eases to $2.02$--$2.08$ below the $m{=}4$ peak while nDCG keeps falling.
The endpoint ($m{=}9$) instead jumps from $2.02$ to $2.58$, positive in all $10$ instances.\footnote{SR averages per-instance ratios, so it need not equal the mean return divided by the mean volatility.}
Here the core shrinks to a single asset, and the falling correlation ($\bar\rho$ $0.11$ to $0.08$) is no longer offset, as individual volatility is now flat ($\times1.00$), so basket volatility drops ($0.228$ to $0.209$) as return rises ($0.48$ to $0.55$) while nDCG collapses to $0.16$.
This endpoint is near-pure diversification, not a re-ranked list.

\begin{table}[t]
\caption{DSER on U92 (curated assets, walk-forward, $k{=}10$).
$m{=}0$ is the expert list and $m{=}9$ the fully decorrelated endpoint. Bold marks the best value for $m{\le}4$.}
\label{tab:u92}
\centering
\small
\renewcommand{\arraystretch}{0.95}
\begin{tabular}{lrrrr}
\toprule
Method & ROI@10 & SR@10 & nDCG@10 & HR@10 \\
\midrule
$m{=}0$ (expert) & 0.242 & 1.794 & \textbf{0.391} & \textbf{0.733} \\
$m{=}1$          & 0.263 & 1.985 & 0.385 & 0.720 \\
$m{=}2$          & \textbf{0.273} & 2.060 & 0.377 & 0.700 \\
$m{=}3$          & 0.266 & 2.085 & 0.370 & 0.685 \\
$m{=}4$          & 0.259 & \textbf{2.119} & 0.362 & 0.668 \\
$m{=}5$          & 0.238 & 2.066 & 0.351 & 0.643 \\
$m{=}6$          & 0.222 & 2.078 & 0.340 & 0.618 \\
$m{=}7$          & 0.200 & 2.046 & 0.315 & 0.566 \\
$m{=}8$          & 0.198 & 2.021 & 0.247 & 0.435 \\
$m{=}9$ (extreme) & 0.239 & 2.575 & 0.160 & 0.284 \\
\bottomrule
\end{tabular}
\end{table}

\paragraph{DSER achieves the best combined relevance--performance score.}
To quantify this balance, we adopt the combined metric of prior work~\cite{pfotgnrec}. 
We min--max normalize both headline metrics to $[0,1]$ over the listed methods, $\widetilde{x}=\allowbreak(x-x_{\min})/\allowbreak(x_{\max}-x_{\min})$ for $x\in\{\mathrm{SR}@10,\allowbreak\ \mathrm{nDCG}@10\}$, and form the $\alpha$-weighted score
\begin{equation}
\label{eq:combined}
\mathrm{Comb}@10(\alpha)\;=\;(1-\alpha)\,\widetilde{\mathrm{nDCG}}\;+\;\alpha\,\widetilde{\mathrm{SR}},
\qquad \alpha\in[0.3,0.7],
\end{equation}
with $\alpha$ weighting portfolio performance ($\widetilde{\mathrm{SR}}$) and swept over the range reported in~\cite{pfotgnrec}.
Because DSER lies strictly inside the range of both metrics, including it does not change the baselines' normalization. The claim is not about a single tuned $m$. Across the entire range, both operating points $m{=}2$ and $m{=}4$ outperform every comparison method (the expert anchor, the prior-art re-ranker RURA, the relevance-led BPR/LightGCN/TGAT, and the Sharpe-led RandomForest/SGL) with no per-$\alpha$ retuning (Table~\ref{tab:combined}).
The leading point shifts only between the two ($m{=}2$ at $\alpha{=}0.3$, $m{=}4$ otherwise).
Both are Pareto-optimal and sit away from the corners of the relevance--performance plane (Figure~\ref{fig:landscape}). 
DSER loses only at those corners, where $m{\ge}5$ yields to high-nDCG BPR at $\alpha{=}0.3$, and where single-objective extremes (higher raw Sharpe, near-zero relevance) lead only outside $[0.3,0.7]$.

\begin{table}[t]
\caption{Combined score $\mathrm{Comb}@10(\alpha)$ (Eq.~\ref{eq:combined}) on U92
(walk-forward, $n{=}10$). Both metrics are min--max normalized over the listed methods
(RURA at its intended risk setting). Best value per $\alpha$ column in \textbf{bold}. Both
$m{=}2$ and $m{=}4$ outperform all baselines across $\alpha\in[0.3,0.7]$.}
\label{tab:combined}
\centering
\small
\setlength{\tabcolsep}{3.4pt}
\resizebox{0.95\columnwidth}{!}{%
\begin{tabular}{lrrrrrrr}
\toprule
Method & SR@10 & nDCG@10 & $\alpha{=}0.3$ & $\alpha{=}0.4$ & $\alpha{=}0.5$ & $\alpha{=}0.6$ & $\alpha{=}0.7$ \\
\midrule
Popularity~\cite{cremonesi2010} & 1.315 & 0.444 & 0.677 & 0.581 & 0.484 & 0.387 & 0.290 \\
MarketAverage~\cite{fartrans} & 2.157 & 0.065 & 0.248 & 0.331 & 0.414 & 0.497 & 0.580 \\
RandomForest~\cite{breiman2001} & 2.332 & 0.088 & 0.342 & 0.436 & 0.530 & 0.624 & 0.718 \\
BPR~\cite{bpr} & 1.616 & 0.442 & 0.762 & 0.695 & 0.629 & 0.562 & 0.495 \\
LightGCN~\cite{lightgcn} & 1.648 & 0.409 & 0.713 & 0.658 & 0.603 & 0.548 & 0.493 \\
SGL~\cite{sgl} & 2.094 & 0.209 & 0.487 & 0.527 & 0.567 & 0.606 & 0.646 \\
MVECF~\cite{mvecf} & 2.054 & 0.206 & 0.470 & 0.506 & 0.543 & 0.580 & 0.616 \\
TGN~\cite{tgn} & 1.900 & 0.151 & 0.327 & 0.363 & 0.398 & 0.434 & 0.469 \\
TGAT~\cite{tgat} & 1.478 & 0.457 & 0.748 & 0.664 & 0.580 & 0.496 & 0.412 \\
PfoTGNRec~\cite{pfotgnrec} & 1.864 & 0.113 & 0.247 & 0.289 & 0.331 & 0.373 & 0.414 \\
Expert ($m{=}0$)~\cite{expertfollow} & 1.794 & 0.391 & 0.723 & 0.687 & 0.651 & 0.615 & 0.579 \\
RURA~\cite{rura2026} & 1.830 & 0.395 & 0.741 & 0.708 & 0.674 & 0.641 & 0.607 \\
\midrule
\textbf{DSER $m{=}2$} & 2.060 & 0.377 & \textbf{0.777} & 0.770 & 0.764 & 0.758 & 0.751 \\
\textbf{DSER $m{=}4$} & 2.119 & 0.362 & 0.768 & \textbf{0.771} & \textbf{0.774} & \textbf{0.778} & \textbf{0.781} \\
\bottomrule
\end{tabular}%
}
\end{table}

\paragraph{Provider reading.}
For a provider, $m{=}2$ is a low-cost operating point and $m{=}4$ favors Sharpe for risk-averse segments.
The relevance cost falls harder on HR than on nDCG, because HR is a binary hit metric structurally sensitive to rarely bought satellites, so Section~\ref{sec:limitations} reads this offline cost as an upper bound.

\subsection{After-Cost Capacity}
\label{sec:capacity}
RQ2 asks whether the gross gain survives trading costs. 
DSER's satellites are comparatively less liquid and hence costlier to trade.
We charge each asset a one-way cost under a square-root market-impact model~\cite{almgren2005},
\begin{equation}
\label{eq:impact}
c_i \;=\; c_{\text{lin}} + c_{\text{imp}}\sqrt{(V/k)/\mathrm{ADV}_i},
\end{equation}
so the basket pays a round-trip $\mathrm{RT}{=}2\,\overline{c_i}$ ($\overline{c_i}$ averaged over the $k$ assets, doubled for entry and exit), and net SR amortizes it over the $T$ holding-window trading days, $\mathrm{net\,SR}{=}\sqrt{252}\,(\overline{r}_B-\mathrm{RT}/T)/\sigma_{r_B}$. Each decision's basket is charged one entry and one exit, so intermediate rebalancing is not priced.
Here $V$ is the per-user portfolio size (swept below), $V/k$ the equal-weight per-asset notional, $\mathrm{ADV}_i$ asset $i$'s average daily traded value (a static proxy, the full-sample average of FAR-Trans brokerage transaction values rather than market-wide volume), and $c_{\text{lin}}{=}10$\,bp with a dimensionless impact coefficient $c_{\text{imp}}{=}0.01$.
Because this proxy is not calibrated to market liquidity, the thresholds below are conditional sensitivity results rather than validated deployment limits. Under it, the net gain survives at retail scale.
At \eur{}$100$k, DSER $m{=}2$ nets $1.64$, above a random fill ($1.60$) and the expert list ($1.56$), its higher gross Sharpe more than offsetting its modestly higher cost.
The advantage is capacity-bounded, however. 
As portfolios grow, the thinner satellites cost more, so $m{=}4$ erodes first and by \eur{}$300$k even $m{=}2$ falls just below the expert list ($1.37$ vs.\ $1.43$).
This is a per-user bound.
With one list for all users, fleet-wide demand concentrates on the same thin satellites, so capacity at scale is lower still.

\section{Ablations: What Drives the Gain?}
\label{sec:ablation}

RQ3 asks what inside DSER produces the gain, so that the provider knows which parts of the design are essential.
We first trace the mechanism. 
We then hold the core, list size, and protocol fixed and vary only the satellite-selection rule and the correlation estimate.

\subsection{The Knob Works by Reducing Crowding}
Because each appended satellite is chosen to be least correlated with the current basket, the basket's realized average pairwise return correlation over the holding window falls about $30\%$ from $m{=}0$ to $m{=}4$ ($0.24$ to $0.17$), yet its realized volatility is flat ($0.225$ to $0.224$).
Under the equal-weight approximation $\sigma_B \approx \bar\sigma\sqrt{1/k + (1-1/k)\bar\rho}$, the correlation drop alone scales basket volatility by $0.89$, but average individual volatility $\bar\sigma$ rises by a factor of $1.12$, a product near $1.00$, so the realized volatility drop is under $1\%$. The approximation is exact only when the assets' volatilities are equal. With our heterogeneous volatilities, it still reproduces the realized basket volatility within $1.3\%$ for $m{\le}4$, and its predicted change from $m{=}0$ to $m{=}4$ ($\times0.993$) matches the realized one ($\times0.993$).
The Sharpe gain therefore flows through the return numerator, the annualized mean return rising about $16\%$ ($0.43$ to $0.50$ in basket aggregates, not the per-instance SR) across $m{=}0$--$4$.
(The buy-and-hold ROI@$k$ of Table~\ref{tab:u92} rises more modestly, $+7\%$.)
Because the satellites are selected by correlation alone, this premium is market-phase dependent (Section~\ref{sec:robust}), whereas the relevance cost is not.

\subsection{What the Selection Rule Contributes}
We isolate the selection rule by filling the same $m$ slots by minimum correlation (DSER), by minimum variance, and at random.
The main gain comes from reducing crowding in the list.
Every breadth-adding rule improves on the crowded expert anchor, but the minimum-variance rule captures the least of it (DSER $+0.24$ SR over minimum-variance at $m{=}4$, $+0.06$ at $m{=}2$, with relevance within $0.01$ nDCG).
The reason is that these low-volatility assets are large, crowded, and correlated with the core, forgoing the return premium that drives the gain. 
This is the ceiling the RURA reimplementation also hits (Figure~\ref{fig:landscape}).
Random filling also reduces crowding, but with high variance (over $30$ seeds, SR $2.03\pm0.10$ at $m{=}4$).
DSER's deterministic $2.12$ ($+0.09$ over the random mean) sits at the $80$th percentile of that distribution.
A sector-based fill uses external sector labels to keep almost all relevance (nDCG $0.39$ vs.\ $0.36$) at a solid Sharpe (SR $2.01$ vs.\ $2.12$ at $m{=}4$), yet DSER leads on the combined score across the performance-weighted operating points ($\alpha{\ge}0.5$) using return correlation alone.

\subsection{Integer $m$ vs.\ Continuous MMR}
\begin{figure}[t]
  \centering
  \includegraphics[width=0.82\linewidth]{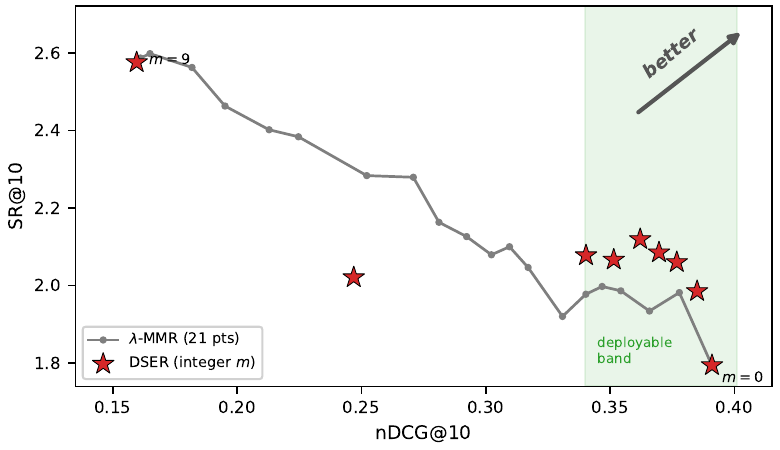}
  \Description{Line plot of SR@10 against nDCG@10 comparing a 21-point lambda-swept MMR curve with DSER's integer-m points. The two coincide at the endpoints, m=0 at nDCG 0.39 and m=9 at nDCG 0.16. Within the shaded deployable band of nDCG at least 0.34, every DSER point lies above the MMR curve.}
  \caption{Integer $m$ vs.\ continuous $\lambda$ (U92, $n{=}10$). DSER (stars)
  and a $\lambda$-swept MMR with the same criterion (line) coincide at the
  endpoints, and within the deployable band (shaded, nDCG$\ge$0.34) DSER lies
  above MMR at matched relevance.}
  \label{fig:mmr}
\end{figure}

\looseness=-1
DSER applies a hard-threshold form of the relevance--diversity trade-off that MMR expresses with a continuous weight.
The relevance term is frozen to the expert core, the content-similarity term is replaced by the full return correlation (off-diagonal structure), and the control is a single integer knob.
We therefore ask whether this hard form sacrifices anything relative to the continuous one.
The comparison is a $\lambda$-weighted MMR with the same expert relevance (Eq.~\ref{eq:expert}) and correlation criterion, scoring each candidate $i$ by $\lambda\,\widetilde{q}_i-(1-\lambda)\,\overline{\rho}_i$, where $\widetilde{q}_i$ is the normalized expert score and $\overline{\rho}_i$ its mean correlation to the items already selected. The first slot is taken by the highest-relevance item. We sweep $\lambda\in[0,1]$ over $21$ points. At the extremes, $\lambda{=}1$ recovers the expert list ($m{=}0$) and $\lambda{=}0$ DSER's most-decorrelated list ($m{=}9$). 
In the high-relevance region a provider uses (nDCG $\ge 0.34$, $m{=}1$--$6$), DSER's integer-$m$ curve lies above $\lambda$-MMR at matched relevance (Figure~\ref{fig:mmr}: $\Delta$SR $+0.08$ to $+0.17$, with $+0.17$ at $m{=}4$, in all $10$ instances). 
MMR leads only in the low-relevance corner (nDCG $<0.30$) that a recommender must avoid.
Because nDCG discounts relevance by position, DSER concentrates its whole relevance cost in the least-valued satellite slots and leaves the top $k-m$ items untouched, whereas $\lambda$-MMR can perturb any slot from the second onward.
The hard, auditable integer form thus costs nothing where it matters.

\subsection{The Gain Is Not an Estimation Artifact}
A correlation-based selector could, in principle, merely exploit estimation error in the sample correlation matrix.
On the curated, well-conditioned U92 it does not. 
Replacing the sample correlation with a Ledoit--Wolf shrinkage estimator~\cite{ledoitwolf2004} leaves the Sharpe gain unchanged ($\Delta$SR $+0.015$, not significant at $m{=}4$), and the gain survives an explicit liquidity filter ($+0.30$ SR over the expert list). 
This is genuine return structure, not a fragile fit to a noisy covariance matrix.

\subsection{Why Experts Are Ranked by ROI}
Ranking experts by raw ROI rather than by a risk-adjusted score might seem mismatched to our Sharpe objective, but the opposite holds. 
Ranking investors by their holdings' Sharpe instead lowers the base list's SR by $0.51$ while raising nDCG and HR by under $0.04$ (Table~\ref{tab:expertsel}). 
Risk-adjusting the selection shrinks and skews the expert pool, whereas acting on the basket (which DSER's decorrelation does) is complementary to selecting experts by ROI.

\begin{table}[t]
\caption{Expert-selection criterion (pure expert list, U92, walk-forward, $n{=}10$, a matched single-process run whose by-ROI SR differs slightly from that in Table~\ref{tab:u92}). Best value per column in \textbf{bold}.}
\label{tab:expertsel}
\centering
\small
\begin{tabular}{lrrrr}
\toprule
Expert selection & ROI@10 & SR@10 & nDCG@10 & HR@10 \\
\midrule
By ROI (ours) & \textbf{0.246} & \textbf{1.831} & 0.390 & 0.733 \\
By Sharpe     & 0.166 & 1.321 & \textbf{0.425} & \textbf{0.749} \\
\bottomrule
\end{tabular}
\end{table}

\subsection{What the Satellites Look Like}
The assets DSER appends are real but thinly traded, with a median daily traded value roughly $1/70$ of the expert core's (\eur{}$315$ vs.\ \eur{}$22$k), and their idiosyncratic price paths are what decorrelate the basket while also raising its trading cost (Section~\ref{sec:capacity}).
The satellite set is nonetheless stable across rebalances. 
Adjacent bi-weekly checkpoints share $0.75$ of their satellites (median, with $35$ distinct assets over the run), so only about a quarter of the satellite positions change at each rebalance (counted by asset, not by traded value).
The gain is not carried by a few assets either. 
Banning the five most-used satellites and re-running does not diminish the Sharpe advantage ($\Delta$SR $+0.53$ at $m{=}4$, if anything, above the full-pool $+0.33$, positive in all $10$ instances).
DSER re-selects other decorrelated substitutes, so the gain reflects low correlation rather than specific names.

\section{Robustness}
\label{sec:robust}

RQ4 asks whether the gain holds under stricter evaluation protocols. 
We stress-test it and state plainly where it holds.

\subsection{Window Overlap}
\looseness=-1
Our walk-forward instances slide in $2$-week steps while the evaluation window is $6$ months, so adjacent windows overlap heavily and the $10$ instances are far from independent. Paired significance is therefore optimistic.
We re-evaluate on non-overlapping $6$-month windows ($n{=}6$), tiled across the sample and spanning different market phases.
The effect is heterogeneous across phases rather than monotone in market stress (Table~\ref{tab:indep}). It is largest in the 2021 rally ($+0.54$ SR), still positive through the 2020 COVID crash ($+0.08$ SR, largest ROI gain $+0.068$), and negative only in the 2020 rebound and the 2022 drawdown, two phases in which the thin satellites lowered risk-adjusted return.
DSER's benefit thus accrues on average rather than in every window.
The SR gain stays directionally positive ($4$ of $6$ windows, mean $+0.15$) but is not statistically significant at $n{=}6$. The $\pm0.26$ spread suggests low power rather than a vanished gain. Detecting a $+0.15$ mean at this dispersion with $80\%$ power would need $n\approx25$ windows (over a decade), beyond any FAR dataset. ROI is positive in $5$ of $6$ windows (median $+0.016$).
The relevance side, by contrast, is directionally consistent across market phases.
Both $\Delta$nDCG and $\Delta$HR are negative in all $6$ windows (reaching $p{=}0.031$, the smallest value attainable at $n{=}6$ under a two-sided sign test), a small, consistent cost of the satellite swap.
The trade-off is thus asymmetric.

\begin{table}[t]
\caption{DSER ($m{=}4$) vs.\ the expert list over $6$ non-overlapping $6$-month windows
spanning distinct market phases (Sharpe gain positive in $4$ of $6$ windows, relevance cost present in all $6$).
In the mean row, the ``$>0$'' counts are the number of windows with a positive change, in column order ROI/SR/nDCG/HR.}
\label{tab:indep}
\centering
\footnotesize
\setlength{\tabcolsep}{3pt}
\begin{tabular}{llrrrr}
\toprule
Window & market phase & $\Delta$ROI@10 & $\Delta$SR@10 & $\Delta$nDCG@10 & $\Delta$HR@10 \\
\midrule
2019\,H2 & late bull    & $+0.016$ & $+0.31$ & $-0.015$ & $-0.040$ \\
2020\,H1 & COVID crash  & $+0.068$ & $+0.08$ & $-0.009$ & $-0.017$ \\
2020\,H2 & rebound      & $-0.125$ & $-0.13$ & $-0.021$ & $-0.034$ \\
2021\,H1 & rally        & $+0.049$ & $+0.54$ & $-0.028$ & $-0.053$ \\
2021\,H2 & late bull    & $+0.009$ & $+0.25$ & $-0.036$ & $-0.106$ \\
2022\,H1 & drawdown     & $+0.015$ & $-0.13$ & $-0.021$ & $-0.058$ \\
\midrule
mean & {\scriptsize ($>$0: 5/4/0/0 of 6)} & $+0.005{\scriptstyle\,\pm0.068}$ & $+0.15{\scriptstyle\,\pm0.26}$ & $-0.022{\scriptstyle\,\pm0.009}$ & $-0.052{\scriptstyle\,\pm0.030}$ \\
\bottomrule
\end{tabular}
\end{table}

\subsection{Single Train/Test Split}
We adopt the prior-work protocol exactly, a single $8{:}1{:}1$ split with one held-out 2022 bear-market test window.
Increasing $m$ still lowers relevance monotonically, and no single method dominates both axes, but the magnitude of DSER's Sharpe advantage attenuates in this single bear window, reversing at $m{=}2$ and not distinguishable from zero at its interior peak ($m{=}4$) under a Jobson--Korkie/Memmel test.

\begin{table}[t]
\caption{Sensitivity of the Sharpe gain to the correlation lookback and the expert share $X$ (U92, walk-forward, $n{=}10$). Each cell is $\Delta$SR@10 of DSER over the expert list with the same $X$. Parentheses give the number of positive instances when fewer than $10$. The paper's setting is $X{=}5$ with a $180$-day lookback.}
\label{tab:sens}
\centering
\footnotesize
\setlength{\tabcolsep}{4pt}
\begin{tabular}{lrrrrrr}
\toprule
 & \multicolumn{2}{c}{$X{=}5$} & \multicolumn{2}{c}{$X{=}10$} & \multicolumn{2}{c}{$X{=}20$} \\
\cmidrule(lr){2-3}\cmidrule(lr){4-5}\cmidrule(lr){6-7}
Lookback & $m{=}2$ & $m{=}4$ & $m{=}2$ & $m{=}4$ & $m{=}2$ & $m{=}4$ \\
\midrule
90 days  & $+0.35$ & $+0.62$ & $+0.21$ & $+0.22$ & $+0.32$ & $+0.59$ \\
180 days & $+0.27$ & $+0.32$ & $+0.20$ & $+0.02$\,(5) & $+0.25$ & $+0.49$ \\
360 days & $+0.21$ & $+0.10$ & $+0.06$\,(6) & $-0.24$\,(0) & $+0.15$ & $+0.25$ \\
\bottomrule
\end{tabular}
\end{table}

\subsection{What Correlation the Rule Needs}
\looseness=-1
The gain is not a stale-price artifact (Section~\ref{sec:related}). 
Re-estimating satellite correlations on weekly or monthly returns, which reduce nonsynchronous-trading bias~\cite{lo2002,getmansky2004}, preserves it ($\Delta$SR $+0.32$ weekly, positive in all $10$ instances, with the monthly gain positive but attenuated).
Table~\ref{tab:sens} sweeps the lookback window and $X$. The gain is larger at shorter lookbacks and stays positive in every instance at $X{=}5$. With $X{=}10$, however, the $m{=}4$ gain vanishes at the default $180$-day lookback and reverses at $360$ days. Recency matters more than sampling frequency.

\section{Limitations}
\label{sec:limitations}
\begin{enumerate}
\item[(1)] We evaluate on a single dataset (FAR-Trans, Section~\ref{sec:setup}). DSER does not fit a predictive model, although its correlation estimates and design choices remain sample-dependent.
\item[(2)] We scope every claim to a curated, liquid universe of exchange-listed equities (the setting prior work adopts) and to the expert-following backbone we test, and do not claim transfer to uncurated or multi-asset universes, where stale prices and illiquidity make correlations unreliable.
\item[(3)] Retention and revenue are proxied by risk-adjusted return, not measured. That return is a basket-level provider measure, not any single user's realized return.
\item[(4)] Our relevance metrics are offline and observational. They score each list against purchases users made without seeing it, so they cannot show whether the re-ranked list would change what users buy, which needs an online test. Because the top $k{-}m$ items are identical to the expert list, the relevance change comes entirely from the $m$ replaced slots. nDCG and HR credit an item only if the user organically bought it, so thinly traded satellites are structurally penalized (a missing-not-at-random bias), making the measured relevance cost plausibly an upper bound.
\item[(5)] The satellite return premium may partly reflect a neglected-stock effect. DSER beats uniform random breadth by $+0.09$ SR at $m{=}4$ (Section~\ref{sec:ablation}), but its minimum-correlation rule also picks thinner names, so this gap does not separate the correlation effect from size and liquidity factors, a separation that would need factor data beyond FAR-Trans.
\end{enumerate}

\section{Conclusion}
\label{sec:conclusion}

The expert-following list that best balances relevance and portfolio performance exhibits internal crowding. 
DSER, a transparent, training-free re-ranker, keeps the expert core and fills $m$ satellite slots with the least-correlated assets, turning the relevance--performance trade-off into a single provider-controlled knob.
Under the primary overlapping-window protocol, $m{=}4$ increases SR from 1.794 to 2.119 while reducing nDCG from 0.391 to 0.362.
The $m{=}2$ gain survives realistic trading costs at retail scale but erodes as portfolios grow.
Across $6$ non-overlapping windows, the Sharpe change is positive in $4$, while the relevance change is negative in all $6$. The knob's cost is thus consistent and its benefit phase-dependent, so DSER gives the provider an explicit, auditable operating point on the relevance--performance trade-off.
Future work includes validation on a second brokerage dataset, extension to multi-asset universes, personalization of the satellites to each user's holdings, and evaluating diversification at the level of each user's resulting portfolio.

\begin{acks}
This work was supported by JSPS KAKENHI Grant Numbers JP24K02942, JP23K11211, and JP24KJ0324.
\end{acks}

\bibliographystyle{ACM-Reference-Format}
\bibliography{refs}

\end{document}